\documentclass[11pt]{article}
\usepackage{euscript}
\usepackage{amsfonts}
\usepackage{sidecap}
\usepackage{hyperref}
\usepackage{url}
\usepackage{mathptmx}
\usepackage{times}
\usepackage{amsmath,amssymb,amsthm,euscript,latexsym}
\usepackage{mathrsfs}
\usepackage{bm}
\usepackage[dvips]{graphicx}

\usepackage{exscale}
\usepackage{color}
\usepackage{tikz}
\usetikzlibrary{calc,patterns,decorations.pathmorphing,decorations.markings}
\usetikzlibrary{snakes,arrows,shapes}
\usepackage{pgf}
\graphicspath{{images/}}

\newcommand{\bmat}{\left[ \begin{matrix}}
\newcommand{\emat}{\end{matrix} \right]}

\newcommand{\beq}{\begin{equation}}
\newcommand{\eeq}{\end{equation}}

\newcommand{\nd}{\noindent}
\newcommand{\q}{\quad}
\newcommand{\qq}{\qquad}

\newcommand{\var} {\mbox{\rm var}}
 
\newcommand{\rank} {\mbox{\rm rank}\,}

\newcommand{\Frac}[2]{{\displaystyle\frac{#1}{#2}}}
\newcommand{\met}{\frac{1}{2}}
\newcommand{\Span} {\mbox{\rm span}\,}

\newcommand{\E} {\mathbb{E}\,}
\newcommand{\Cbb} {\mathbb{C}}
\newcommand{\Rbb} {\mathbb{R}}

\newcommand{\Nbb} {\mathbb{N}}

\newcommand{\Hbb} {\mathbb{H}}
\newcommand{\Zbb} {\mathbb{Z}}

\def\v{{\rm v}}
\def\ib{{\mathbf i}}

\def\bb{{\mathbf b}}
\def\ab{{\mathbf a}}
\def\hb{{\mathbf h}}
\def\cb{{\mathbf c}}

\def\gb{{\mathbf g}}
 \def\qb{{\mathbf q}}
\def\pb{{\mathbf p}}

\def\ub{{\mathbf u}}

\def\xb{{\mathbf x}}
\def\vb{{\mathbf v}}
\def\yb{{\mathbf y}}
\def\wb{{\mathbf w}}
\def\zb{{\mathbf z}}

\def\Cb{{\mathbf C}}

\def\Hb{{\mathbf H}}

\def\Xb{{\mathbf X}}
\def\Ub{{\mathbf U}}

\def\deltab{\boldsymbol{\delta}}
\def\xib{\boldsymbol{\xi}}

\def\pib{\boldsymbol{\pi}}
\def\mub{\boldsymbol{\mu}}

\def\etab{\boldsymbol{\eta}}

\def\lambdab{\boldsymbol{\lambda}}

\def\Sigmab{\boldsymbol{\Sigma}}

\newtheorem{thm}{\bf Theorem}

\newtheorem{lem}{\bf Lemma}
\newtheorem{cor}{\bf Corollary}

\newtheorem{prop}{\bf Proposition}

\begin{document}

\thispagestyle{empty}

\begin{center} 
{\textcolor{blue}{\Huge {  On Irreversibility\\[4. mm] and Stochastic Systems\\ [4. mm] Part Two}}} \\[1.cm]

 in memoriam of  Jan. C. Willems\\[.5 cm]

{\Large   GIORGIO PICCI}\\[.6cm]
Department of Information Engineering\\
  University of Padova, Italy\\
  \tt{picci@dei.unipd.it}

\end{center}

\nd{\bf Abstract:} We attempt to characterize irreversibility of a dynamical system from the existence of different forward and backward mathematical representations depending on the direction of the time arrow. Such different representations have been studied intensively and are shown to exist  for stochastic diffusion models. In this setting one has however to face the preliminary justification of the existence of a stochastic description for physical systems which are traditionally described by classical mechanics as inherently deterministic and conservative.\\
 In part one of this paper we have  addressed this modeling problem from a deterministic viewpoint for linear systems. We have shown that there are forward-backward representations which can describe conservative finite dimensional deterministic systems when they are  coupled to an infinite-dimensional conservative heat bath. A key observation is that   the heat bath acts on the finite-dimensional system by {\em state-feedback}  which can shift its eigenvalues to make the system dissipative, but also may generate a totally unstable system which naturally evolves backward in time. 
 
 In this second part, we  address the stochastic description of these two representations. Under a natural family of invariant measures  it will be shown that the heat bath  induces  a white noise input acting on the forward-backward representations  making them true dissipative diffusions.

\newpage

\section{\bf The Stochastic Picture}\label{StochChap}
One may speculate that  all physical or engineering  systems obey first-principle physical theories which lead to describe them by   {\em conservative} (or lossless)  models, as for example  theoretical mechanics. Models of this kind are always deterministic \cite{Willems-03}. But real systems by their very nature are always coupled to an external environment which absorbs energy and gives rise to an observable dissipative behavior. In this paper we imagine to model  this external environment  as an infinite dimensional Hamiltonian system which plays the role of a  {\em  heat bath} in Physics.  Macroscopic  models usually describe the coupling to the environment      as an external drag or  resistance acting on the system but these models are of an empirical nature and do not   take faithfully into account the fine details of this coupling. For example a microscopic model of an electric circuit should try to describe the fluctuations of the electrons in the conductor which is the origin of dissipation. This at the end  implies that you may better describe the system by a ``noisy" {\em stochastic model}. \\
   Any such stochastic model,   by its very nature, turns then out to  be {\em irreversible}: it must in fact  imply   different stochastic realizations depending on the direction of time.

\subsection{\bf Program of this chapter:} 
To discuss the following issues
\begin{description}
\item [1] What is the analog of the Maxwell-Boltzmann distribution on the phase space of an infinite-dimensional Hamiltonian system, say on $\ell^2_{\Cbb}$?
\item [2] With this distribution the motion of a particle  will  become a stationary stochastic process. When is this  motion  described by a  {\bf purely non deterministic (ergodic) process} and how can we link it to the deterministic dynamics? Under these conditions we can prove the $H$-theorem.

\item [3] When is  there  a finite-dimensional  stochastic dynamical model representing  the motion of this process, in fact as a function of a  p.n.d (or ergodic) Markov process. This will lead to generalized {\em Langevin equation} representation of the process.
\end{description}

In the following it will be necessary to distinguish sharply between variables belonging to  an infinite dimensional (separable) Hilbert space, which will be denoted by {\bf bodface} symbols and their coordinates with respect to some basis (i.e. arrays of real or complex numbers) which will be denoted by usual lowercase.

\subsection{\bf Invariant probability distributions on the phase space $\Hb$}
 
 To fix notation for future use,  it is convenient to use the complex change of variables already used in some of the examples. We shall assume that the overall linear Hamiltonian system  is defined  by  the coupling of  a finite dimensional lossless linear system with  a linear infinite-dimensional Hamiltonian heat bath defined on a Hilbert phase space  $\Hb$ by a positive densely defined potential operator admitting a symmetric square root $V$. We can then make the time-evolution of the system  {\em unitary} by introducing a complex (infinite-dimensional) phase vector $z(t): = p(t)+jV q(t)$  ($j= \sqrt(-1)$) which leads  the system to evolve in time on a complex Hilbert space, which for concreteness will  be identified with $L_{\Cbb}^2(\Rbb_+) $.   This makes the Hamiltonian function look like the square norm of the phase  variable namely $H(z)=\met \|z\|^2 $ and the evolution of the heat bath system, taking place  in the  complex $L_{\Cbb}^2(\Rbb_+)$ space,   preserves the   norm of $z$. The canonical equations can be rewritten in a skew-symmetric form as
$$
\frac{d}{dt}\bmat p(t) \\ jVq(t)\emat = j \bmat  0& V\\ V &0\emat   \bmat p(t)\\ jVq(t)\emat \qquad \equiv \q
\dot z(t)= jAz(t)  ; \qquad A^{*}=A
$$
So that $ jA=-[ jA]^*$   is {\bf skew adjoint}  and, under appropriate technical conditions which we shall not spell out here, the steady-state evolution   taking   place in the (complex) Hilbert phase space,  is governed by the relation
$$ 
z(t) = e^{ j At }z(0) := \Phi(t) z(0)\,;\qquad t\in \Rbb
$$
 where the $\{ \Phi(t);\,t\in \Rbb\} $  is a {\em continuous  group of unitary linear operators on the phase space}.\\ 
 Let   $\Xb$ be the finite-dimensional state space of a given lossless system coupled  to the heat bath. Naturally $\Xb$ can be identified with  a finite-dimensional subspace of a larger Hilbert pace, say $\Xb \vee L_{\Cbb}^2(\Rbb_+) $. For the example of the electrical line (without the lossless load), the   time evolution of the heat bath takes place on the complex $L_{\Cbb}^2 (\Rbb_+)$ space or, equivalently, on the  direct sum $\Hb = L^2(\Rbb_+) \oplus L^2(\Rbb_+)$,  here $\{ \Phi(t);\,t\in \Rbb\} $   is   just translation (of course still unitary). For the particle system  the Hamiltonian phase space becomes   $\ell_{\Cbb}^2(\Zbb)$ and the Hamiltonian is again one half of the norm of $z$ (an infinite square-summable complex  sequence).

\nd In this chapter we shall mostly concentrate on the analysis of the infinite-dimensional Hamiltonian  heat bath. It is not hard to check that the spectrum of $A$, denoted  $\sigma \{ A \}$,  is the union of the spectrum of $V$ and that of $-V$. Since  $V$ is the positive self-adjoint square root  of $V^2$, the spectrum of $A$  is the union of two opposite real components symmetric  w.r. to the origin of the real line. Therefore 
    $$
 \sigma \{ j A \}\,\subset  \,\text{The Imaginary axis} \, 
 $$
 which is the starting point for the spectral theory reviewed in Sect. \ref{SpectralTheory} where a fundamental result is that, in the present setting,  $\Phi(t)$ must be {\em  unitarily equivalent to the translation operator} $\Sigma(t)$ on a suitable $L^2(\Rbb)$ space.  From this equivalence valid for the complexified Hamiltonian picture, we can then  derive the existence of a continuous group of time translations, which we may still denote by  $\Phi(t)$, which describes the time evolution of a real linear Hamiltonian system. See e.g. formulas (18) and (19) in Part one.

\nd{\em On Invariant probability distributions on the phase space} $\Hb$.\\
Recall that we have proved using Liouville theorem that in a finite--dimensional phase space, any
absolutely continuous $\Phi (t)$-invariant probability measure admits a density
$\rho (p,q)$  equal to a normalization  constant times 
$exp[-\frac{1}{\beta}H(p,q)] , \beta:= kT >0 $ where $H(p,q)$ is the Hamiltonian function. For a quadratic Hamiltonian function we could actually identify this density with a Gaussian and $\beta$ with the variance  $ \sigma^2=kT$.\\
We would now  like   to find a $\Phi(t)$-invariant measure $\mu$  which is  {\em countably additive} on the Hilbert phase space $\Hb$ of an infinite dimensional Hamiltonian system and thereby make the phase variables $\{\pb(t),\qb(t); t\in \Rbb\}$ evolving on the Hilbert space $\Hb$ into bona-fide   stochastic  processes  defined on a probability space where $\Omega\equiv \Hb$, i.e. say $\{\Hb, \mathcal A, \mu\}$. \\
It is necessary to warn the reader that there is a technical difficulty  with {\em continuous--spectrum} Hamiltonian
systems in a infinite dimensional phase  space $\Hb$,  that, although  there are natural invariant absolutely continuous
  probability measures,  they  in general    {\em cannot be extended as a countably additive measure} to the whole space $\Hb$.  A survey of the difficulties inherent in  this problem is for example provided   in \cite[Sect.3]{Hida-80}. However,   as already  indicated in \cite{Picci-T-92}, in our special context we need to describe  only some finite-dimensional observables of the system and there is  no need to look for countably additive measures on the whole space.

 \subsection{\bf Cylinders and multiplicity}
 Let $\cb_1,\ldots,\cb_m$ be fixed vectors in a  Hilbert space $\Hb$. A {\it Cylinder} is a subset of $\Hb$ defined by
\beq \label{Cyl}
\Cb_{\cb_1,\ldots,\cb_m}:=  \{z \mid \langle \cb_1,z\rangle \ldots  \langle \cb_m,z\rangle\;\in B \},\quad B\in {\mathcal B}_m
\eeq
 where  ${\mathcal B}_m$ is the sigma-algebra of Borel subsets of $\Cbb^m$. Clearly the pre-images through  linear functionals of the Borel sigma-algebra ${\mathcal B}_m$ form a sigma-algebra of subsets of $\Hb$, say ${\mathfrak C}_{\cb_1,\ldots,\cb_m}$, which we call the {\em sigma-algebra of cylinder sets with basis $\cb_1,\ldots,\cb_m$}.\\
  Now let $\Phi(t), t\in \Rbb$ be a strongly continuous  group of not necessarily unitary, operators on $\Hb$,   and consider the time-indexed family of cylinders
 \beq \label{Tcyl}
 \Cb_{\cb_1,\ldots,\cb_m}(t):=  \{z \mid \langle \cb_1,\Phi^*(t)z\rangle \ldots  \langle \cb_m,\Phi^*(t)z\rangle \;\in B \},\quad B\in {\mathcal B}_m\,. 
\eeq

 \begin{lem}\label{LemCyl}
 The sigma-algebra generated by by the time-indexed family of cylinders \eqref{Tcyl} for all $t\in \Rbb$,  coincides with ${\mathfrak C}_{\cb_1,\ldots,\cb_m}$. 
There is always a    countably additive extension of any $\Phi(t)$-invariant measure  to {\bf cylinder spaces}, that is to the sigma-algebra of cylinder sets with a finite  basis $\cb_1,\ldots,\cb_m$.
 \end{lem}
 \begin{proof}
 Since for each fixed $t$, $\Phi(t)$ is  one-to-one and onto, the image of each  $z\in \Hb$  through the measurable map $z^{\prime}:=\Phi(t)^*z$  belongs to  a cylinder of the form \eqref{Cyl} which can be represented by the same relation with the variable $z$  renamed $z^{\prime}$. Hence for each fixed $t$ this cylinder belongs to ${\mathfrak C}_{\cb_1,\ldots,\cb_m}$. The same is of course true for the inverse mapping and hence the two sigma-algebras are identical for each fixed $t$. Next since $\Phi(t)$ is continuous we may as well restrict $t$ to the countable subset of rational numbers and hence the sigma-algebra  generated by all ${\mathfrak C}_{\cb_1,\ldots,\cb_m}(t)$ for all rational $t$'s coincides with ${\mathfrak C}_{\cb_1,\ldots,\cb_m}$.  \end{proof}
 
 Note that the proof does not require $\Phi(t)$ to be unitary, in other words the Hamiltonian does not need to be quadratic and reducible to the  norm in Hilbert space as it happened in some of the previous examples. See also \cite[Def. 3.4]{Hida-80}.

\subsection{\bf The Maxwell-Boltzmann distribution on the phase space $\Hb$}

 Assume a linear infinite-dimensional Hamiltonian of the form $H(\pb,\qb)= \|\pb\|^2 +\met\langle \qb, V^2 \qb\rangle$ on some separable Hilbert phase space $\Hb$. Let $\{\hb_k, k\in \Nbb\}$ be a complete orthonormal  system in $\Hb$ and let $\Hb_N$ be the subspace spanned by the first $N$ elements of the basis. The first $N$ coordinates with respect to this basis, of the canonical variables $\pb(t)$ and $\qb(t)$ say
\beq  \label{Coordinates}
q_k(t):= \langle \hb_k,\qb(t)\rangle, \qq p_k(t):= \langle \hb_k,\pb(t)\rangle,\q k=1,2,\ldots,N
\eeq
written as $N$-column vectors, say $q_N(t), p_N(t)$, must satisfy the finite-dimensional canonical equations
$$\begin{cases}
\bmat \dot q_N(t)\\ \dot p_N(t)\emat= \bmat 0  & I\\ - V_N^2 & 0\emat \,\bmat q_N(t) \\p_N(t) \emat
\end{cases}
$$
so that the quadratic Hamiltonian $H(\pb,\qb)= \|\pb\|^2 +\met\langle \qb, V^2 \qb\rangle$, once restricted to the finite-dimensional subspace $\Hb_N$, takes the form
\beq \label{FinHamilt}
H(p_N,q_N)= \|p_N\|^2 +\met\langle q_N, V_N^2 q_N\rangle
\eeq
where the norm and the inner products are Euclidean and $V_N^2$ is  a positive definite symmetric matrix.\\
Now the group of linear operators $\{\Phi_N(t),t\in \Rbb\}$ describing the Hamiltonian flow of the reduced system, acts on  the coordinates of $(p,q)$ in  $ \Rbb^{2N}$ (we shall not use a different symbol for it) and, by construction,   the Hamiltonian function is $\Phi_N-$ invariant, i.e.
\beq  \label{TimeinvH}
H(p_N(t),q_N(t)):= H(\Phi_N(t)\bmat p_N\\q_N\emat)= H(p_N,q_N) \,, \qq t\in \Rbb
\eeq
From what we have seen at the beginning of this story, there is a $\Phi_N(t)$-invariant measure  on the finite dimensional subspace $\Hb_N$ having a  density  of the Maxwell-Boltzmann type:
 \beq \label{FinDens}
 \rho_{N}(p_N,q_N)= C_N\,\exp \{ -\frac{1}{2\beta} H(p_N,q_N) \},\qq (p_N,q_N)\in \Rbb^N\times \Rbb^N
 \eeq
We shall show that this measure can be extended to a countably additive measure on any cylinder space $\{\Hb, {\mathfrak C}\}$ having finite basis.

\subsection{\bf Characteristic functions}

 Recall the definition of Characteristic function for  n-vector random variables $\xb(\omega)$ and real n-dimensional variable $\lambda^{\top}:= [\lambda_1,\ldots,\lambda_n]$: 
 $$
 C_{\xb}(\lambda): = \int_{\Omega} \exp \{j\,\lambda^{\top} \!\xb(\omega)\} P(d\omega)= \int_{\Rbb^n}\exp {j\langle \lambda,\,x \rangle}\, dF_n(x)\,
 $$
 which is generalized to Hilbert spaces with probability measure $\mu$ and Hilbert space variable $\lambdab$ as 
 $$
 C_{\xb}(\lambdab): = \int_{\Hb} \exp {j\langle \lambdab,\,\xb\rangle}\, d\mu(x)\,
 $$
 It is well-known that there is a one-to-one correspondence between measures (or random variables)  and their characteristic functions.
 \begin{thm}[Levy-Gnedenko-Helly Bray]\label{LevyThm}
 Let $\{\xb_k, \, k=1,2,\ldots\} \subset \Hb$. If and only if $C_{\xb_k}(\lambdab)$ converges pointwise at each  $\lambdab$ then the corresponding distributions $\mu_{\xb_k}$ converge weakly to a limit probability distribution.
 \end{thm}
 From this basic result one can prove existence of an extension of the Maxwell-Bolztmann measures on $\Hb$ by a limit argument.
 \begin{thm}
The measures $\mu_N$ of density  \eqref{FinDens}
 converge  weakly as $N\to \infty$ to a Gaussian measure on $\{\Hb, {\mathfrak C}\}$ having characteristic function
 \beq \label{LimitCharact}
 C(\pib,\xib)= \exp \{ -\frac{ 1}{2}\sigma^2 H(\pib, \xib) \}, \qq (\pib,\xib)\in \Hb\times \Hb.
 \eeq
 \end{thm}

\begin{proof}
Assume for concreteness that $\pb,\qb$ are functions in $L^2(\Rbb_+)\times L^2(\Rbb_+) $ and each function $\hb_k$ in the ON basis has two components $\hb_k= [\pib_k, \xib_k]$ forming mutually orthonormal sequences. The characteristic function of the density \eqref{FinDens} is
\begin{align*}
C_N(\pib,\xib) &= \int_{\Rbb^{2N}} \exp   j \{ \sum_{k=1}^{N} \langle \pib_k, \pb\rangle +\langle \xib_k, \qb\rangle\}\, d \rho_{N}(p_N,q_N)\\
&=\exp -\frac{ 1}{2}\sigma^2\bmat \pib_N\\ \xib_N\emat^{\top} \bmat I_N &0\\0 &V_N^2\emat \bmat\pib_N\\ \xib_N\emat  \\
& = \exp-\frac{ 1}{2}\sigma^2 \{ \|\pib_N\|^2 +\langle \xi_N,V_N^2 \xi_N\rangle \}
\end{align*}
where $\pib_N$ and $\xib_N$ are $N$-vectors with components $\pib_k$ and $ \xib_k$ and $V^2_N$ is the $N\times N$ matrix representation of the operator $V^2$ with respect to the   basis $\xib_N$.  The sum in the exponent of the last member  converges pointwise as $N\to \infty$ to $ \exp \{ -\frac{ 1}{2}\sigma^2 H(\pib, \xib) \}$ for any such orthonormal sequence $\{\hb_k\}$.      Hence by  Theorem \ref{LevyThm} the density \eqref{FinDens} must converge weakly to a Gaussian  probability measure which has zero mean and quadratic  covariance operator $\sigma^2\bmat I&0\\0 &V^2\emat$, a positive densely defined operator on $L^2(\Rbb_+)\times L^2(\Rbb_+) $. In case we need to interpret $\langle \pib_k, \pb\rangle$ and $\langle \xib_k, \qb\rangle$ as distributions, we may interpret  them as linear functionals of  ``smooth" test functions $\pb, \qb$. See \cite[Chap.~3]{Hida-80}.
\end{proof}
  
   \begin{cor}
 There are $\Phi(t)$-invariant absolutely continuous measures $\mu_{\beta}$  which are  countably additive on any cylinder sigma-algebra ${\mathfrak C}$ of $\Hb$  having a finite basis $\cb_1,\ldots,\cb_m$.
 \end{cor}
 It is in fact known that  when ${\mathfrak C}$ is a cylinder sigma-algebra with a finite basis the measure $\mu_{\beta}$ will be countably additive on the probability space $\{ \Hb, {\mathfrak C}, \mu_{\beta}\}$ \cite[Sect. 3.2]{Hida-80}.

\subsection{\bf Generalized stochastic processes and White Noise}
 
For the examples that we have seen so far the Hamiltonian of the heath bath has the standard structure $H(p,q)= \met \{\|p\|^2 +\langle q,V^2q\rangle\}$ where the canonical variables are functions in $L^2(\Rbb_+)$ (or in $\ell^2_{\Cbb}$)  and $V^2$ a positive, densely defined self-adjoint operator. It is  obvious, from   the multiplicative structure of the invariant measure,   that the $\pb$ and $\qb$ processes are {\bf independent} Gaussian processes and the marginal density of the $\pb$ random variable (in fact process)   must have the form
\beq\label{Distrp}
\mu_{p}= Const. \times \exp {-\frac{1}{2\beta} \|\pb\|^2}
\eeq
which is that of a {\em white noise process}, see \cite[Chap. 3]{Hida-80}. In fact, we shall show that

\begin{thm}\label{pWhite}
The random variable $\pb= \{ \pb(x) \,;\, x\in \Rbb_+\}$, once understood as a  generalized stochastic process (see \cite[Chap. 3]{Hida-80}) with trajectories in $L^2(\Rbb_+)$, has independent sample values $\pb(x)$ and is , in fact,  a  {\bf white noise} process. 
\end{thm} 
\begin{proof}
We should in principle try to show that for any $x\neq y$ in $\Rbb_+$, the random variables  $\pb(x)$ and $\pb(y)$ are independent. Of course there are several difficulties to carry on this argument. First, obviously enough,  the sample value $p(x)$ of a function in  $L^2(\Rbb_+)$ is undefined and, second, even if $p(x)$ was a continuous function, the independence of arbitrarily close sample value cannot hold as the variable $\pb(x)$ should then have zero variance and be deterministic.\\
Therefore the argument needs to be framed in the setting of generalized stochastic process. To this purpose we shall resort to the Gel'fand Rigged Hilbert Spaces embedding \cite{Gelfand-V-64}
$$
{\mathcal S}\subset \Hb\subset{\mathcal S}^{\prime}
$$
where ${\mathcal S}$ is a Hilbert space of smooth functions densely imbedded in $\Hb$, which for $\Hb\equiv L^2(\Rbb_+)$ could be a Sobolev space of absolutely continuous functions and ${\mathcal S}^{\prime}$ its dual, containing linear functionals  such as the `` delta functions" denoted $\deltab_x$ which act on smooth functions as
\beq
\langle \deltab_x, f\rangle= f(x) \qq \text{for } \qq f\in {\mathcal S}\,.
\eeq
A technical fact which we shall be giving for known, is that the  $\deltab_x$ functionals can be approximated by smooth test functions like a shrinking Gaussian function of the variable $\xi$, centered at the point $x$, say  $\gamma_{\epsilon}(x): \xi \to C \exp\{-\met \frac{(\xi-x)^2}{\epsilon^2}\}$, in the sense that
\beq
\lim_{\epsilon\to 0}\,\langle \gamma_{\epsilon}(x), f\rangle = f(x) \qq \text{for } \qq f\in {\mathcal S}\,.
\eeq
Now, because of  \eqref{Distrp} and its covariance structure,  the correlation of two smooth random samples $\pb(x)$ and $\pb(y)$ in ${\mathcal S}$, equal to
$$
  \E\{ \langle \deltab_x,\pb\rangle, \langle\deltab_y,\pb \rangle\}= \beta \langle \deltab_x,\delta_y\rangle_{{\mathcal S}^{\prime}}
= \lim_{\epsilon\to 0}\,\beta\langle \gamma_{\epsilon}(x), \gamma_{\epsilon}(y)\rangle_{L^2}
$$
must be equal to zero whenever  $x\neq y$ since the two approximating functions tend to be orthogonal.  Hence the argument must hold in a generalized sense   by a limit approximation of the two delta functionals.
\end{proof}

 Therefore, 
 \begin{cor}\label{WhiteIC}
 Under the invariant measure $\mu_{\beta}$,  the initial values  of the voltage and  current of the infinite electrical line, $\{\v_0(x); x\in \Rbb_+\}$ and $\{\, i_0(x); x\in \Rbb_+\}$,  become independent random processes and both of them are white noise of variance $\sigma^2$. The corresponding initial values at $t=0$ of the waves derivative processes $a^{\prime}(x),\,\text{\rm and}\,\,b^{\prime}(x); \, x\in \Rbb_+$ defined in formulas (16) and (17) (for $t=0$) of Part One  are also white generalized processes.
 \end{cor}
 \begin{proof}
 In fact, because of the standardization of the $L,C$ parameters both $\v_0$ and $i_0$ satisfy the same wave equation and hence the same Hamiltonian equation (4) of Sect. 3.2 of Part one where they both can play the role of the momentum $(p)$ variable. Therefore Theorem \ref{pWhite} implies that both are white noise. Next since (formally) $i_0(x)= \frac{1}{C}\dot {\v}_0(x),\, x\in \Rbb_+$,  is uncorrelated with $\v_0(x)$ by the multiplicative structure of the invariant measure, and both processes are Gaussian, they must be independent  white noise processes.\\
 Hence (16) and (17)   imply that both $\ab^{\prime}=\{a^{\prime}(x)\,; x\in \Rbb_+\}$ and $\bb^{\prime}:= \{b^{\prime}(x)\,; x\in \Rbb_+\}$  are also white processes.
 \end{proof}

\begin{lem}
Under the invariant measure $\mu_{\beta}$ the time-shifted processes $\ab_t^{\prime}$ and $\ab_s^{\prime}$ are stationary  white noise generalized processes taking values in  $L^2(\Rbb_+)$. They are  independent whenever  $t\neq s$ and both have  second moment $\beta^2 \|\ab_0^{\prime}\|^2$ the norm being that of $L^2(\Rbb_+)$. A completely analogous statement holds for $\bb_t^{\prime}$ and $\bb_s^{\prime}$. 
\end{lem}
 \begin{proof}
 Since $\ab_0^{\prime}:=\met (\vb_0 +\ib_0)$ with  $\vb_0$ and $\ib_0$ independent  white noises both of variance $\beta^2$, it follows that $\ab_0^{\prime}$ is also a white noise of variance $\beta^2$. The same is true for $\ab_t^{\prime}= \Sigma_t\ab_0^{\prime}$,
 (compare (18) of Part one)  for arbitrary $t$. Therefore if $t\neq s$ the correlations of all random sample values $\ab_t^{\prime}(x)$ and $\ab_s^{\prime}(y)$ must be zero, i.e.
 \beq\label{WNa}
 \E \ab_t^{\prime}(x)\ab_s^{\prime}(y)=0,\qq \forall x,y \in \Rbb
\eeq
  while for $t=s$ one has $\E [\ab_t^{\prime}(x)]^2= \beta^2\|\ab_0^{\prime}\|^2$, since $\Sigma(t)$ is unitary.
 \end{proof}
 Let us now consider the time-indexed processes $\{w(t)\equiv a^{\prime}(t); \,t\in \Rbb\}$. Since \eqref{WNa} can be written
 $$
 \E  a^{\prime}(t+x)a^{\prime}(s+x)= \beta^2\|\ab_0^{\prime}\|^2 \delta(t+x-(s+x))=\beta^2\|\ab_0^{\prime}\|^2 \delta(t-s)
 $$
for all $x\in \Rbb_+$,  it follows that in particular $w(t)=a^{\prime}(t)=a^{\prime}(t+x)_{x=0};\, t\in \Rbb$ must be a white process for all times and the same must hold true for   $\{\bar{w}(t) \equiv b^{\prime}(-t); \,t\in \Rbb\}$.
 
 In conclusion, under the invariant measure for the Hamiltonian heat bath, described in Part one,  for the electrical line example, by equation (4) or by equation (41) for the mechanical string, both stochastic process inputs $w,\bar w$ to the deterministic Langevin-like equations (34) and (35),  are Gaussian white noise. By analogy this also happens for the mechanical  example of Sect. (3.5). Hence these equations become bona-fide {\bf stochastic differential equations} describing the motion of the state variable of an arbitrary linear observable, say of the form (33) or (45). As anticipated in Remark (9) of Part one, an analogous statement can be shown to hold for the motion of the Brownian particle. It will be argued in the next section \ref{StatProc}   that these observable processes are all {\em purely non deterministic} and in fact,  {\bf ergodic}. For them the $H$-Theorem holds and hence {\em these systems are irreversible also according to the classical entropy-based definition of  irreversibility}.\\
  So far, however,  we have only been  discussing particular physical instances. In the next section we discuss this {\em irreversible representation problem} in more generality.

 \section{\bf The  Hilbert space theory of stationary processes}\label{StatProc}
Under the invariant measure,  an arbitrary linear observable of an  infinite-dimensional Hamiltonian system  evolves in time as a   {\em stationary Gaussian process}. The statement being obviously true also for a finite number of such linear observables which would then  lead to vector-valued processes.

These objects can be completely analyzed in a (stochastic) Hilbert space  setting which, to some extent, avoids the necessity of considering generalized stochastic processes.  The Hilbert space  theory of stationary processes  is a large body of results  whose foundations are  due to A.N. Kolmogorov, H. Wold, H. Cram\`er, K. Karhunen and others. It is exposed in many places but the most complete treatment of the foundations seems still to be classical book by Rozanov \cite{Rozanov-67}. In this section we shall mostly refer to this book and to the subsequent literature.

One very important class of such processes are the {\em purely non-deterministic (p.n.d.) ones} \footnote{Called {\em regular} by \cite[p. 178]{Rozanov-67}.  A related concept for deterministic dynamical systems, not necessarily linear, is (intuitively)  that they should  generate regular signals. These are called K-systems, after Kolmogorov.}. It is a well-known fundamental result that   under a very weak continuity condition \footnote{Automatically satisfied for the linear processes that we are considering.}{\bf   purely non deterministic stationary processes are ergodic} see \cite{Parzen-58},  \cite[Chap. IV, p. 178]{Rozanov-67}. Hence for these processes the tentative proof of the H-theorem in Sect. 2.4 of Part One is in fact valid.  \\
This brigs us back to question 2 of the Program of this Part Two as sketched in the beginning.

 \subsection{\bf Stochastic Hilbert spaces and  Spectral theory}\label{SpectralTheory}

In this subsection we shall again  interpret $\Hb$ as the phase space of a linear Hamiltonian system with   $\zb$   denoting the  conjugate variables $\bmat\pb&\qb\emat^{\top}$ written in complex form.The  space of linear functionals on a Hamiltonian Hilbert phase space $\Hb$  once equipped with an invariant measure $\mu_{\beta}$, can be  identified with  a ``stochastic" Hilbert space $\Hbb$ whose elements are random variables which are the  linear functionals on $\Hb$. Under an invariant measure all these random variables  have a  Gaussian distribution. The correlation of, say, $\xb(\zb)= \langle \xb, \zb\rangle$, $\hb(\zb)= \langle \hb, \zb\rangle; \q \xb,\hb,  \zb \in \Hb$,  is 
\beq  \label{Isometry}
\E\xb\bar \hb:= \langle \xb,\hb\rangle_{\Hbb}:=  \langle    \xb, \Sigma\, \hb\rangle_{\Hb}\,
\eeq
where
 \beq
\Sigmab:= \sigma^2\bmat I&0\\0& V^2\emat
\eeq
denotes the covariance operator of the measure $\mu_{\beta}$. If $\Sigma$ is invertible then
$\Hbb$ and $\Hb$ are of course  isometric.\\
We shall lift the Hamiltonian group $\Phi(t)$  to act on these  random variables   by letting
  \beq \label{Ulift}
  \Ub(t)\hb:\, \zb\to \,\langle  \cb,  \Phi(t) \zb\rangle, \qquad t\in \Rbb,\; \zb\in \Hb
\eeq
which defines  $\Ub(t)$ as a {\em one-parameter  group  acting on random variables} which  is continuous in m.s. since the  group $\Phi(t)$ is strongly continuous on $\Hb$. In fact, $\Ub(t)$ is unitarily equivalent to the adjoint $\Phi(t)^{*}$ acting on linear functionals of $\zb\in \Hb$.\\
 Suppose one can observe the motion of the Hamiltonian system along some fixed vector $\cb \in \Hb$; then,    the scalar  process 
$$
\yb(t)=  \langle \cb,\, \zb(t)\rangle := \Ub(t) \yb(0), \qquad  \yb(0):=\langle \cb,\, \zb(0)\rangle,\qquad   t\in \Rbb
$$ 
is a {\em stationary m.s. continuous  Gaussian process}. In general, for  $m$ observation functionals
$\langle \cb_1,\zb\rangle \ldots  \langle \cb_m,\zb\rangle $ we can define a stationary  $m$-dimensional Gaussian stochastic process on $\{ \Hb, {\mathfrak C}_{\cb_1,\ldots,\cb_m}, \mub\}$ by letting:
\beq \label{VectProc}
\yb(t, \zb):= \bmat \langle \cb_1, \Phi(t)^{*}\zb\rangle \\ \ldots \\ \langle \cb_m, \Phi(t)^{*}\zb\rangle \emat =\bmat \Ub(t)\langle \cb_1, \zb\rangle \\\ldots \\ \Ub(t)\langle \cb_m, \zb\rangle \emat
\eeq
 In what follows we shall only be interested in  functions (statistics) of  vector processes of the form \eqref{VectProc}. Since this process   is measurable with respect to the cylinder sigma algebra ${\mathfrak C}_{\cb_1,\ldots,\cb_m}$ (compare Lemma \ref{LemCyl}), all these functions will also  be measurable with respect to this cylinder sigma-algebra. Therefore we shall  not need to worry about extensions of the measure $\mu_{\beta}$ to a larger space.

 Let $jA$, with $A^{*}=A$,  be  a skew adjoint operator (which we assume unbounded with dense domain) on an arbitrary separable Hilbert space $\Hb$; it is well known \cite{Dunford-Schwartz-64}  that it has a spectral representation
$$
jA= \int_{-\infty}^{+\infty} j \lambda  \, dE(j\lambda)\,
$$
where $E$ is an {\em orthogonal projection-valued measure}: For any Borel subsets $\Delta_1, \Delta_2$ of the imaginary axis and $\hb,\gb \in \Hb$,
\begin{align}
 \text{if}\q \Delta_1\cap\Delta_2 = \emptyset & \Rightarrow \q E(\Delta_1\cup \Delta_2)= E(\Delta_1)+ E(\Delta_2)\\
E(\Delta_1)E(\Delta_2) &=   E(\Delta_1 \cap\Delta_2)\\
 \langle E(\Delta_1)\hb, E(\Delta_2)\gb\rangle &=  \langle \hb, E(\Delta_1 \cap\Delta_2)\gb\rangle = \mu_{\hb,\gb}(\Delta_1 \cap \Delta_2),
 \end{align}
where $\mu_{\hb,\gb}$ is a complex measure which can depend on $\hb,\gb$. Any such skew adjoint operator generates a strongly continuous unitary group and {\bf conversely} any strongly continuous unitary group $\{\Phi(t);\, t\in \Rbb\}$ on the Hilbert space $\Hb$ has the representation:
\beq
\Phi(t)= \int_{-\infty}^{+\infty} e^{j\lambda t} dE(j\lambda) \equiv  e^{jA t}\,,\qq t\in \Rbb
\eeq
  for some skew adjoint operator $jA$,  this is the {\em Stone-von Neumann} theorem.

  \nd{\em Finite Multiplicity:} There can be many such measures $\mu$, in general as many as the cardinality of an orthonormal basis but in our setting we can   restrict to unitary groups of {\it finite multiplicity}  for which we have
    $$
 \Hb= \Hb_0:=\Span_{\{t\in \Rbb\} } \{ \Phi(t) \cb_k\,; k=1,\ldots,m\}
 $$
  For any such finite set of generators $\{\cb_k\}$   we can define the {\em spectral Matrix measure on the imaginary   axis } see e.g  \cite[p. 105-115]{Fuhrmann-81} as
  \beq
  M(\Delta):= \left[ \mu_{\cb_k,\cb_j}(\Delta)\right]_{k,j=1 \ldots,m}
\eeq
and the corresponding {\em Spectral distribution matrix}
\beq
F(\lambda):= M([-\infty, j\lambda])
\eeq
which is nonnegative definite, monotonically increasing (in matrix sense) but not necessarily bounded above.
 When $M$ is equivalent (i.e. mutually absolutely continuous) to {\it Lebesgue measure} we say that the operator $jA$ has {\em Lebesgue spectrum}.

\subsection{\bf  Stochastic processes on linear spaces}

 Consider the continuous  unitary group $\Ub(t)$  acting on   random variables $\hb(z)\in \Hbb$, in particular  acting on a scalar random variable $\yb(0):= \langle \cb,z\rangle $. This action generates a stationary stochastic process $\{\yb(t)\}$ and  the spectral representation of unitary groups directly yields the  {\em spectral representation of the stationary process} $\{\yb(t)\}$:
  $$
  \yb(t)=\Ub(t) \yb(0)= \int_{-\infty}^{+\infty} e^{j \lambda t} \, d\hat E(j\lambda)\yb(0) =\int_{-\infty}^{+\infty} e^{j \lambda t} \,  d \hat{\yb}(j\lambda)   
$$
where   the spectral projection  acts on linear functionals as $\hat E(\Delta)\langle \hb,\zb\rangle:= \langle E(\Delta) \hb, \zb\rangle_{\Hb}$ so that
$$
d \hat{\yb}(j\lambda)= d\hat E(j\lambda)\yb(0) 
$$
which  is  a {\em stochastic orthogonal measure}, the {\bf Fourier transform} of $\{\yb(t)\}$.  This is the most direct and economical way to define the  Fourier transform of a stationary process,  due to Kolmogorov; see \cite[p. 16-17]{Rozanov-67}.

  The stochastic measure $\hat\yb$ induces a positive Borel measure $\mu_{\yb}$ on the imaginary axis such that
$$
\E \langle \hat{\yb}(\Delta_1),\hat{\yb}(\Delta_2)\rangle = \mu_{\yb}(\Delta_1 \cap \Delta_2); 
$$
which  is also written symbolically as
$$
\E |d\hat{\yb}(j\lambda)|^2= d\mu(\lambda)\,.
$$
\nd{\em Definition:   The Spectral Distribution Function (PDF) of the process $\yb$} is a function defined by the position
$$
F_{\yb}(\lambda):= \mu_{\yb}[-\infty, j\lambda]; \qquad \lambda \in \Rbb
$$
which  is  clearly nonnegative  and monotone increasing.\\
When $\mu_{\yb}$ is absolutely continuous, it has the {\bf Spectral Density Function}
$$
\hat \Phi(\lambda):= \Frac {d F(\lambda)}{d\lambda}\,.
$$
All of this can directly  be generalized to vector processes $\yb(0)^{\top}:= \bmat \langle \cb_1,z\rangle&, \ldots,&\langle \cb_m,z\rangle\emat^{\top} $ to define a $m\times m$  matrix PDF. These are usual concepts  in probability theory which we shall give for granted.
\medskip
 
 We shall also need to consider  {\em stationary increments processes} whose spectral representation is  described in detail in \cite[pag. 88-89]{LPBook}.These objects are only defined modulo an arbitrary additive random variable. An important example is the {\bf Wiener Process} $\wb$ which has in fact {\em orthogonal} stationary increments and may be vector-valued. For such an $m$-dimensional Wiener process we have
 \beq 
 \E \|\wb(t)-\wb(s)\|^2= \Sigma \, |t-s|,\qq t,s \in \Rbb
 \eeq
   where $\Sigma$ is an $m\times m$ symmetric positive-definite variance matrix.
 \medskip
 
\nd{\bf Example: The ``Brownian" particle in $\ell^2$}
 
If the measure $\mu_{\yb}$ is finite then
\beq
\mu_{\yb}[-\infty, +\infty]= \beta\|\cb\|^2= \var\{\yb(0)\} = \var\{\yb(t)\} 
\eeq
but in general this may not be the case. When $\mu$ is  Lebesgue measure  one has instead
$$
\E \langle \hat{\yb}(d\lambda),\hat{\yb}(d\lambda)\rangle = \E\|\yb(0)\|^2  d\lambda
$$
and one can prove  \cite[p. 88]{LPBook} that the process $\yb(t)$ has {\bf orthogonal stationary  increments}, in fact 
$$
\E \|\yb(t)-\yb(s)\|^2= \sigma^2 |t-s|
$$
where $\sigma^2= \beta \|\cb||^2$. Since it is Gaussian it is a {\bf Wiener process}. \\
 NB: since the variance of $\yb(t)$  grows linearly without bound one should only consider increments.
 
What kind of Hamiltonian system generates this process?

Suppose $V^2_{k,h}=\delta_{k,h}$, no interaction among particles! Then
$$
H(q,p)= \met \{ \sum_{k}p_k^2 +   \sum_{k}q_k^{2} \}
$$
and the invariant density becomes an infinte product of Gaussian terms 
$$
 \Pi_{k\in \Zbb}\, C_k \exp \{-\frac{1}{2\beta} (p_k^2 + q_k^{2}) \}
 $$ 
Hence: the motion of each particle is {\bf independent of the others} (no interaction) and the infinitesimal generator of $\Phi(t)$ becomes  [CHECK]
 $$
 A= \bmat 0 & I\\ I & 0\emat
 $$
therefore  $E(j\lambda)$ becomes the trivial resolution of the identity: $E(j\lambda) h$ is multiplication by the characteristic function $I_{[-\infty, j\lambda]}$. Hence $\mu_{\beta}$ becomes {\bf Lebesgue measure}.\\
 {\bf Conclusion:} if all particles move independently, the observation  of  their motion along any direction $\cb$, written formally as
 $$
 \yb(t)= U(t) \langle \cb,\,\cdot \rangle 
 $$
is a Wiener process. This is in particular what happens to the displacement variable $q_0(t)$ in the Brownian particle example in Subsect. 3.6 of Part one.

\section{\bf The Markovian Representation Problem (Stochastic Realization)}\label{StochReal}
In part one of this paper we have seen examples where  the output of a linear conservative  dynamical  system  connected to an infinite dimensional Hamiltonian heat bath, can, under an invariant measure for its time evolution, become a continuous stationary Gaussian process. A general  question then arises when this process could be  ergodic (i.e. p.n.d.) and, in fact,  be described by a Langevin-type stochastic differential  equation.

We shall first deal with theses questions from a probabilistic point of view. A dynamical characterization will be discussed later.\\
Let us first note that, since a process described by a stochastic  Langevin equation is just the simplest kind of  Markovian diffusion process, the underlying basic question should be posed in more general terms as follows.

 \nd{\it Question 1}: Let $\yb(t)$ be an $m$-dimensional stationary continuous stochastic process. Is there  a {\bf finite dimensional} jointly stationary,  continuous  Markov process   $\xb(t) := \bmat\xb_1(t) &\xb_2(t)&\dots &\xb_n(t)\emat^{\top}$ and a $m\times n$ matrix $C$ such that $\yb(0)= C \xb(0)$ or, equivalently
$$
\yb(t)= C \xb(t) \qquad t\in \Rbb\,.
$$
{\it Question 2:} When is $\xb(t)$ a diffusion process? That is when does $\xb(t)$ obey a (linear) stochastic differential equation of the type ({\it a vector  ``Langevin Equation"})
\beq \label{LinMarkov}
d\xb(t)=A\xb(t) dt +B d \wb(t),
\eeq
with $\Re e \lambda(A) <0 $ and $\{ \wb(t)\}$  a (normalized) Wiener process ?

This is a {\it Stochastic Realization Problem}. A general answer has been obtained by past research work mostly done jointly by A. Lindquist and G. Picci, condensed in the book \cite{LPBook}. The following theorem is a condensed version of a concrete algorithmic solution of the  problems. \footnote{Notation: The Fourier frequency variable is now denoted $\lambda$  since the symbol $\omega$ is already takemn.}

 \begin{thm} {[L-P]}\label{MainStoch}
There are finite--dimensional Markovian representations of a stationary continuous process $\{\yb(t)\}$ if
and only if the spectral distribution matrix $F$ of the process is absolutely
continuous with a {\bf rational spectral density} $\hat{\Phi}(j\lambda)$,
\begin{equation}
\hat{\Phi}(j\lambda) = \frac{d}{d\lambda}F(j\lambda)
\end{equation}  

Finite dimensional Markovian representations of $\yb(t)$ are obtained by
the following procedure
\begin{enumerate}
\item
Do spectral factorization of $\hat{\Phi}(j\lambda)$, i.e. find a rational
$m\times r$ matrix functons $W$ satisfying
\begin{equation}
\hat{\Phi}(j\lambda) = W(j\lambda)  W(j\lambda)^*  \label{SF}
\end{equation}
 such that $W$ extends to an analytic matrix function on the right-half complex
plane (such a factor  is called {\em analytic}). We restrict for simplicity to
left-invertible factors, in which case $r = \rank(\hat{\Phi}), a.e.$  
 
\item
For each analytic spectral factor $W$ define the Gaussian stationary--increments
process $\{\wb(t)\}$ by assigning its spectral measure $d\hat{\wb}(j\lambda)$ as
\begin{equation}
d\hat{\wb}(j\lambda) := W^{-L}(j\lambda) d\hat{\yb}(j\lambda) \label{dw}
\end{equation}
the superscript $-L$ denoting left inverse. Then $\{\wb(t)\}$ is a
$r$--dimensional vector Wiener process and $\{\yb(t)\}$ has the following spectral
representation \begin{equation}
\yb(t) = \int_{-\infty}^{+\infty} e^{j\lambda t}W(j\lambda)d\hat{\wb}(j\lambda),
\end{equation}
\noindent in terms of $\{\wb(t)\}$.

\item
Find a {\em minimal realization} of $W(j\lambda)$, i.e. compute constant real
matrix triples $\{F,G,H\}$ with $F$ square $n\times n$, $B$ of dimension $n\times r$
and $H$ of dimension $m\times n$ with $n$ as small as
possible, such that
\begin{equation}
W(j\lambda) = H(j\lambda I - F)^{-1}G
\end{equation}
\end{enumerate}
Then, corresponding to such analytic spectral factor $W$, $\{\yb(t)\}$ admits a {\bf Forward Markovian
representation} of the form 
\begin{eqnarray}
  d\xb(t) & = & F\xb(t)dt + Gd\wb(t)             \label{XG}                     \\
   \yb(t) & = & H\xb(t)                        \label{YG}
\end{eqnarray}
where $\Re e \lambda(F)<0$, the representation corresponding to each such $W$ being unique modulo change of basis on the
state space and $r\times r$ constant orthogonal transformations on the Wiener process $\wb$.

\item Reverse time direction: there are {\em co-analytic} spectral factors $\bar W$ which extend  to an analytic matrix function on the left-half complex plane which are in a 1:1 correspondence with the analytic factors. Correspondingly,  $\{\yb(t)\}$ admits a companion {\bf Backward Markovian representation} of the same dimension of \eqref{XG}, having the form 
\begin{eqnarray}
  d{\xb}(t) & = & \bar{F}{\xb}(t)dt + \bar{G}d \bar{\wb}(t)             \label{XGbar}                     \\
   \yb(t) & = & \bar H {\xb}(t)                        \label{YGbar}
\end{eqnarray}
where the state process can be chosen the same as in \eqref{XG}, the matrix $\bar F$ is anti-stable, in fact, the spectrum of $\bar F$ is exactly the opposite of that of $F$, in particular $\Re e \lambda(\bar F)>0$. Now $\bar \wb(t)$ is a {\bf backward Wiener process} and the representation  is unique modulo a change of basis in the same sense as before. The transfer function  between the forward and backward Wiener processes is 
\begin{equation}\label{StochK}
d\hat{\bar{\wb}}(j\lambda)  := K(j\lambda)d\hat{\wb}(j\lambda)
\end{equation}
where $K(j\lambda):= \bar{W}(j\lambda)^{-L}(j\lambda)$ is a unitary matrix function which extends to the complex plane as  an  matrix  inner function, called the {\em Structural  Function} associated to the pair in \cite[p.317]{LPBook}.
\end{thm}
There are more general mathematical results,  see \cite{Lindquist-P-85b}, \cite{Protter-87}, stating that  diffusion processes, which are  special examples of {\em Semimartingales} and, in particular, martingales (like a Wiener process), always have  forward and backward representations which are  {\em different}. They  are intrinsically  {\bf irreversible processes}.  
 
In particular, Wiener processes $\wb$ always come  in pairs of {\bf forward and backward realizations} like $\wb$ and $\bar \wb$  in \eqref{XG}, \eqref{XGbar}, which generate the evolution of any  linear diffusion  process, in the two opposite directions of time,   which are in fact   {\it different !}.

Gaussianness is  an extra bonus coming from the physical way the process may be generated. The   statements of the theorem remain valid also in a wide-sense context without any mention of Gaussian distribution.\\
The dissipative character of both   {\bf Forward and Backward Markovian Semigroups} follows from the fact that $e^{Ft}$ and $e^{{\bar F}t}$ are  contracting in opposite direction of time since the eigenvalues of $F$ have negative real part while those of $\bar F$ are all with positive real part. This  greatly  generalizes the negative coefficient of the momentum in the right hand side of the original (forward)  Langevin's equation \cite[eq. (3)]{Langevin-908}.

\subsection{From Markovian realizations  to coupled lossless systems\\
(The link Hamiltonian-Stochastic)}
In this subsection we face the background question of the paper: how to relate the  stochastics to the  Hamiltonian deterministic picture discussed in Part one. In our current context, 
 can a continuous stationary process   be always thought of, or  represented, as the output of a lossless deterministic system coupled to an infinite dimensional heat bath?

In the current framework, this question has an essentially positive answer. We refer to Theorem \ref{MainStoch}  assuming that the process has a rational spectral density  which means that it can be represented as the output of a forward (and backward) Langevin-type diffusion model \eqref{XG}, \eqref{YG} with an $n$-dimensional  Markov state process $\xb(t)$.
\begin{thm}\label{MainRep}
Any   stationary process with a rational spectral density   can be  represented as the 
 output of a  lossless linear system with a rational  transfer function $Z_0(s)$, coupled to a hyperbolic heat bath in thermal equilibrium. This transfer function is related to the inner function $K(s)$ of formula \eqref{StochK} by the relation
 \beq
 K(s)=(Z_0(s)-I)(Z_0(s)+I)^{-1}\,
\eeq
which establishes a one-to-one correspondence between $K(s)$ and $Z_0(s)$.
\end{thm}
\begin{proof}
We shall first give a proof for a scalar process. The extension to the multivariable setting is sketched at the end.\\
 We want to express the stationary process described by \eqref{YG} as output of a filter with transfer function $Z_0(s)$ driven by some other stationary  input process $\ub$. The pair $(\yb,\,\ub)$ must then be measurable (i.e. function of) the stochastic waves $\ab^{\prime},  \bb^{\prime}$ generated by the heat bath. Because of linearity and the wave structure this can be expressed by imposing the scattering relations 
 \beq \label{StochScatt}
 \ab^{\prime}=\met(\ub +\yb),\q  \bb^{\prime}=\met(\ub -\yb)
 \eeq
 from which we get $ \ub=-\yb +2\ab^{\prime}$ and $\ub= \yb- 2 \bb^{\prime}$ which, using the assumed structure \eqref{YG}  of $\yb$ can be rewritten as
 \begin{align}
  \ub&=-H\xb +2\ab^{\prime}, \q\text{that is}\q  Z_0(s)\hat\ub +\hat \ub =2\hat{\ab}^{\prime}\\
  \ub&=H\xb -2\bb^{\prime}, \q\text{that is}\q  -Z_0(s)\hat\ub +\hat \ub =-2\hat{\bb}^{\prime}
  \end{align}
  where the hats stand for Fourier transforms. The relations on the right are in turn equivalent to
   \begin{align}
 \hat \ub &=(I+ Z_0)^{-1}2\hat{\ab}^{\prime}\\
  \hat \ub&=(Z_0-I)^{-1} 2\hat{\bb}^{\prime}\,.
  \end{align}
  Now recalling that $\hat{\ab}^{\prime}=d\hat{\wb}$ and $\hat{\bb}^{\prime}=d\hat{\bar\wb}$
  and the definition \eqref{StochK} of the function $K$, eliminating $\ub$ we get the equation
  \beq \label{Zzero}
  K=\Frac{1-Z_0}{1+Z_0},
 \eeq 
which is analogous of the relation (30) in Part One. Let  $Z_0(s)= N(s)/D(s)$ be a coprime rational description of $Z_0$ and let $P(s)$ be a real  polynomial with zeros on the left-hand plane so that
\beq
K(s)= P(-s)/P(s)
\eeq
provides  a rational representation of the inner function $K$. Then \eqref{Zzero} provides the relation
\beq
\Frac{N}{D}= \Frac{P(-s) +P(s)}{P(s)-P(-s)}
\eeq
 which, after expressing $P(s)$ as the sum of its  odd and even parts, say $P(s)=P_o(s) +P_e(s)$ where $P_o(-s)= -P_o(s)$ and  $P_e(-s)= P_e(s)$, is seen to be the ratio of an even and odd polynomials, a well-know classical characterization of lossless  (conservative) rational transfer functions \cite {Foster-24}.
 
 \nd{\em Extension to the multivariable case (sketch).}
 The generalization should  involve multivariable wave signals, that is multivariable white noise processes generated by   the heat bath.  Assuming   their dimension $m$ to be the same as that of the output process \eqref{YGbar} ($\yb$ of full rank), one has to work with a rational $m\times m$ matrix inner function $K(s)$. The scalar argument can be generalized  by resorting (with proper care) to coprime {\em Matrix Fraction Description} representations of the relevant matrix transfer functions. We shall skip the details. 
 \end{proof}
Next we point out that, also in the present setting,  the    inner function $K(s)$ of the process only depends on  the  dynamics of the Markovian state and is independent of the particular output signal which need not necessarily be stationary but more generally have stationary increments.
 \begin{prop}\label{InvarLem}
Consider any process $\etab(t)$ obtained as a linear instantaneous function of the  Markov process $\xb(t)$ and forward noise $\wb(t)$, of the type\footnote{Such a process is really not stationary but has stationary increments.}
\beq\label{StatIncrProc}
\etab(t)= C\xb(t) +D \wb(t)
\eeq
which is described by the  transfer function $ T(j\lambda):= C(j\lambda I - F)^{-1}G +D$ (assumed invertible for simplicity)
 and the backward  representation of the same process by a linear function of the same Markov process $\xb(t)$ but backward noise $\bar{\wb}(t)$, that is
 \beq \label{StaIncr}
 \etab(t)= \bar C\xb(t) +\bar D \bar{\wb}(t)
 \eeq
 having transfer function  $ \bar{T}(j\lambda):= \bar C (j\lambda I - \bar F)^{-1}\bar G +\bar D$. Then it holds that $ \bar{T}(j\lambda)^{-l}T(j\lambda)$ extends to the complex plane as  an  inner matrix function, which coincides with the {\em Scattering  Inner Function} of the process $K(s)$, namely
  \beq
\bar{T}(j\lambda)^{-1}(j\lambda)=K(j\lambda)\,.
 \eeq
 Therefore the scattering function $K$ is invariant with respect to the choice of the linear output  functional   $\etab$.
 \end{prop}
 \begin{proof}
 Need first to express the rational transfer functions of  the maps $\wb\to \etab$ and $\bar\wb\to \etab$ based on  the two state  dynamics  \eqref{XG} and \eqref{XGbar} as polynomial  matrix fraction descriptions. Then the  invariance can be proved by polynomial matrix manipulations.
 \end{proof}
  This invariance property seems to be little known but can be seen as  the background of the coordinate-free geometric structure  studied in Chap 8 of the book \cite{LPBook}. It implies that  the representation result of Theorem \ref{MainRep} also holds for any observable stationary increments process having a finite-dimensional description based on a fixed scattering function $K$, as the structure \eqref{StaIncr}. 
  
 Theorem \ref{MainRep} may have a conceptually wider significance: it may be interpreted as an explanation of the {\em origin (or a mechanical explanation)  of randomness } which can be seen, at least for stationary phenomena, as the effect of the interaction of a deterministic {\em finite-dimensional lossless system} with an infinite-dimensional Hamiltonian environment (the heat bath). Stationarity in particular occurs  in thermal equilibrium.\\
There is a curious and seemingly profound analogy with another phenomenon in theoretical  engineering, the {\em Darlington synthesis representation  of positive real functions} \cite{Darlington-84}.This representation  of linear electrical system with dissipation, may be seen in a similar light. The termination of the lossless network on a resistive load just needs to be interpreted as a termination on a an infinitely long lossless electrical line.

\subsection{\bf Conclusions}

We have discussed the  characterization of  irreversibility  it in terms of the {\em mathematical  dynamic model} of a linear physical system both   by  following the empirical dictum:  {\em Irreversible systems must be described by models which are sensitive to the orientation of the time line} and also by a rigorous investigation of system structures for which  the classical condition of strict increase of the entropy is guaranteed. \footnote{{\em ``Orientation"} is a well-defined mathematical concept, well-known  in geometry.
The time line has two different  orientations: one for increasing $t$ the other for decreasing $t$. We stress that  that change of orientation does   not correspond to the {\em reflection with respect to the origin}: $t\to -t$.}\\
 In this respect  the key  observation is that,  while deterministic ODE are invariant with respect to orientation, {\em  stochastic diffusion models are not}. Hence stochastic diffusion models are natural for capturing irreversibility. One may say that 
 stochastic diffusion models of an observable process are intrinsically {\it  irreversible} since any diffusion process  always admits forward/backward pairs of dynamic representation (stochastic realizations)  which  are different.
 In other words any forward (stochastic diffusion) representation  and, in particular its dynamic parameters, must {\em change} if you change direction of time. This appears to be  the {\em mathematical essence} of {\em irreversibility}.


\begin{thebibliography}{99}

\bibitem{Anderson-V-73} B. D.O. Anderson and S. Vongpanitlerd, {\em Network Analysis and Synthesis : a modern Systems Theory approach}, Prentice Hall (1973) reprinted by Dover in 2006.

\bibitem{Boltzmann-895} L. Boltzmann, On Certain Questions of the Theory of Gases, {\em Nature}, vol 51, 413 (1895)

\bibitem{Boltzmann-897}  L. Boltzmann, On Zermelo's Paper "On the Mechanical Explanation of Irreversible Processes"
[Originally published under the title: "Zu Hrn. Zermelo's Abhandlung Über die mechanische Erklärung irreversibler Vorgange", {\em Annalen der Physik}, vol.  60, pp. 392–8 (1897)].

\bibitem{Darlington-84} Darlington, Sidney "A history of network synthesis and filter theory for circuits composed of resistors, inductors, and capacitors", {\em IEEE Transactions: Circuits and Systems}, vol. 31, pp. 3–13, 1984.


\bibitem{Davey-11} 
Davey K. Thermodynamic Entropy and Its Relation to Probability in Classical Mechanics. {\em Philosophy of Science};78(5):955-975, 2011. doi:10.1086/662559

\bibitem{Devaney-89} R.L. Devaney: {\em An introduction to chaotic dynamical
systems}, second edition, Addison Wesley, New York, 1989.


\bibitem{Dunford-Schwartz-64} N. Dunford and J. T. Schwartz {\em Linear Operators Part II}, Wiley Interscience 1963.

\bibitem{Foster-24} Foster, R M, "A reactance theorem",{\em Bell System Technical Journal}, Vol. 3, pp. 259–267, 1924.

\bibitem{Fuhrmann-81} P.A. Fuhrmann {\em Linear Systems and Operators in Hilbert Spaces},
Mc Graw-Hill, 1980.

 

\bibitem{Ford-K-M-65} G.W.Ford, M.Kac, P.Mazur: Statistical mechanics of assemblies of
coupled oscillators, {\em Journal Math. Phys.} {\bf 6}, pp 504-515, 1965.

\bibitem{Gelfand-V-64} I.M Gel'fand and N. Ya Vilenkin, {\em Generalized Functions, Vol 4 : Applications of Harmonic Analysis},
Academic Press, 1964.

\bibitem{Godefroy-S-91} G. Godefroy and J.H. Shapiro: Operators with dense, invariant, 
cyclic vector manifolds, {\em J. Funct. Analysis}, 1991.

\bibitem{Goldstein-80} H. Goldstein {\em Classical Mechanics, 2nd edition}, Addison Wesley, 1980.

\bibitem{Halmos-57} P. Halmos {\em Introduction to Hilbert Space and the Theory of Spectral Multiplicity, second ed.} Chelsea, N.York 1957.

\bibitem{Hida-80} T. Hida: {\em Brownian Motion} Springer, New York,~1980.

\bibitem{Herrero-92} D. Herrero: Hypercyclic operators and chaos, {\em Journal of Operator Theory}, Vol. 28, No. 1 (Summer 1992), pp. 93-103.
\bibitem{Jaynes-65} Jaynes, E. T.  “Gibbs vs Boltzmann Entropies.” {\em American Journal of Physics}  33:391–98, 1965..

\bibitem{Kuo-75} H.H. Kuo: {\em Gaussian measures in Banach spaces}, Springer
Verlag, New York, 1975.

\bibitem{Langevin-908} P.  Langevin, Sur la th\`eorie du mouvement Brownien. {\em Comptes rendus de l'acad\`emie des
sciences de Paris}, pp. 530-533  (1908).

\bibitem{Lax-P-67} P.D. Lax and R.S. Phillips: {\em Scattering Theory}, Ac. Press, New
York, 1967

\bibitem{Lebowitz-93-99}
J.L. Lebowitz, Boltzmann's Entropy and Time's Arrow, {\em Physics Today}, 46, 32–38(1993); see also letters to the editor and response in {\em Physics Today}, 47, 113-116 (1994);{c)} Microscopic Origins of Irreversible Macroscopic Behavior, {\em Physica A}, 263, 516–527, (1999); 

\bibitem{Lebowitz-99b}	J.L. Lebowitz,  A Century of Statistical Mechanics: A Selective Review of Two Central Issues, {\em Reviews of Modern Physics}, 71, 346–357, 1999; {e)} From Time-symmetric Microscopic Dynamics to Time-asymmetric Macroscopic Behavior: An Overview,  in {\em European Mathematical Publishing House}, ESI Lecture Notes in Mathematics and Physics. 

\bibitem{Lebowitz-08} Joel L. Lebowitz , Time's Arrow and Boltzmann's Entropy, {\em Scholarpedia}, 3(4):3448.	doi:10.4249/scholarpedia (2008). 



\bibitem{Lewis-T-75} J.T.Lewis and L.C.Thomas: How to make a heat bath, in {\em
Functional Integration and its Applications}, A.M.Arthurs ed, p. 97-123,  Oxford, Clarendon,
1975.

\bibitem{Lewis-M-84} J.T.Lewis and H. Maassen: Hamiltonian models of classical and
quantum stochastic processes, in {\em Spriger Lect. Notes in Math.} n.{\bf
1055}, pp 245-276, 1984.


\bibitem{Lindquist-P-79} A Lindquist and G. Picci, On the Stochastic Realization Problem {\em SIAM Journal on Control and Optimiz.}, {\bf
17}, pp. 365-389, (1979). doi.org/10.1137/03170.


\bibitem{Lindquist-P-85} A Lindquist and G. Picci, Realization Theory  for multivariate
stationary Gaussian processes {\em SIAM Journal on Control and Optimiz.}, {\bf
23}, pp1-50, 1985.

\bibitem{Lindquist-P-85b}  A Lindquist and G. Picci, Forward and Backward semimartingale representations of stationary increments processes, {\em Stochastics}, Vol. {\bf 15}, No. 5 ,pp. 1-50, 1985. 

\bibitem{Lindquist-P-91} A Lindquist and G. Picci, A geometric approach to modeling and
estimation of linear stochastic systems {\em Journal of Math. Syst. Estimation
and Control}, {\bf 1},pp. 241-333,1991.

\bibitem{LPBook} A Lindquist and G. Picci,{\em  Linear Stochastic Systems: A geometric approach to modeling 
estimation and identificationof linear stochastic system}, Springer Verlag 2015.

\bibitem{Massen-90} H. Maassen : Hamiltonian models of classical and quantum stochastic
processes  in {\em Realization and modeling in System Theory}, pp 505-511,
Birkhauser Boston, 1990.

\bibitem{MAA-thesis} H. Maassen, On a class of quantum Langevin equations and
the question of approach to equilibrium. {\em Thesis}, University of Groningen,
1982.


\bibitem{Nelson-58} E. Nelson, The adjoint Markoff process, {\em  Duke Mathematical Journal}, vol. 25 (1958), 671–690.

\bibitem{Nelson-67}  E.  Nelson, {\em  Dynamical Theories of Brownian Motion}. Princeton University Press (1967).\\
url:https://web.math.princeton.edu/~nelson/books/bmotion.pdf 

\bibitem{Parzen-58} Emanuel Paŕzen, Conditions That a Stochastic Process be Ergodic
{\em The Annals of Mathematical Statistics}, Vol. 29, No. 1 (Mar., 1958), pp. 299-301 

\bibitem{Picci-86} G. Picci: Application of stochastic realization theory to a
fundamental problem of statistical physics, in {\em Modeling, Identification and
Robust Control}, C.I.Byrnes and A.Lindquist eds., North Holland,1986.

\bibitem{Picci-88} 	G. Picci: Hamiltonian representation of stationary processes
	in {\em Contributions to Operator Theory and its applications}   I. Gohberg,
J.W. Helton, L. Rodman eds.,pp.193-215,  Birkhauser, Boston, 1988.


\bibitem{Picci-88bis} 	G. Picci:
	Stochastic aggregation, in {\em Linear Circuits, Systems and Signal Processing}  C.I.  Byrnes, C.
Martin, R. Saeks eds.  (Proceedings of  the 8th International Symposium on the
Mathematical Theory  of Networks and Systems, Phoenix, Arizona), North Holland, pp.493-501, 1988.

\bibitem{Picci-89} G. Picci, Aggregation of linear systems in a completely
deterministic framework, in {\em Three decades of Mathematical Systems Theory: A
Collection of Surveys at the occasion of the Fiftieth Birthday of Jan C.
Willems}, H. Neijmeijer, J.M. Schumacher eds. {\em Springer Lect. Notes in
Control and Information Sciences}, {\bf 135}, pp 358-381, 1989.


\bibitem{Picci-T-90} G. Picci and T.J. Taylor, Stochastic aggregation of linear
Hamiltonian systems with microcanonical distribution, in {\em Realization and
modeling in System Theory}, pp 513-520, Birkhauser Boston, 1990.


\bibitem{Picci-91}  G. Picci: Markovian representation of linear Hamiltonian systems, in
{\em  Probabilistic Methods in Mathematical Physics},
Western Periodicals, Singapore, 1991.


\bibitem{Picci-92}  G.Picci: Stochastic model reduction by aggregation in {\em
Systems Models and Feedback: Theory and Applicatons}, pp. 169-177, Birkhauser
Boston, 1992.


\bibitem{Picci-T-92} G. Picci and T. J. Taylor: Generation of Gaussian Processes and Linear Chaos, in
{\em Proceedings of the 31st Conf. on Decision and Control} IEEE Press, Tucson, AZ, 1992, pp 2125-2131. 




\bibitem{Prigogine-80} I. Prigogine, {\em From Being to Becoming}, W. H. Freeman, San Francisco, 1980.

\bibitem{Protter-87}  P.   Protter: Reversing Gaussian semimartingales without Gauss. {\em Stochastics} 20; 39-49, (1987).

\bibitem{Rozanov-67} Y.A. Rozanov, {\em Stationary Random Processes}, Holden Days, S.
Francisco, 1967.

\bibitem{Schrodinger-51} E. Schr\"odinger, Irreversibility,
{\em Proceedings of the Royal Irish Academy} vol.53, 1951.

\bibitem{Steckline-83} V.S. Steckline, Zermelo, Boltzmann and the recurrence paradox. {\em Am. J. Physics}, vol. 51 (1983) 894-897.

\bibitem{Willems-72} J.C. Willems,  Dissipative dynamical systems part I: General theory. {\em Arch. Rational Mech. Anal.} 45, 321–351 (1972). https://doi.org/10.1007/BF00276493

\bibitem{Willems-03} Jan C. Willems:  Reflections on Fourteen Cryptic Issues Concerning the Nature of Statistical Inference, {\em International Statistical Review}, Vol. 71, Issue 2, (Aug 2003) , pp. 277-318.

\bibitem{Willems-86}  Jan C. Willems: From time series to linear system—Part I. Finite dimensional linear time invariant systems, {\em Automatica} 1986, p. 567.

\bibitem{Zermelo-896} E. Zermelo, Ann. Physik 57, 485 (1896). 

\bibitem{Zermelo-896b} E. Zermelo, Ann. Physik 59, 793 (1896).
 


\end{thebibliography}
\end{document}